\documentstyle[12pt]{article}

\begin{document}
\baselineskip=14pt plus 0.2pt minus 0.2pt
\lineskip=14pt plus 0.2pt minus 0.2pt
\begin{center}
\Large{\bf Coherent States and Squeezed States, \linebreak
Supercoherent States and Supersqueezed States}
\\

\vspace*{0.25in}

Michael Martin Nieto

\vspace*{0.1in}

Theoretical Division, Los Alamos National Laboratory\\
University of California\\
Los Alamos, New Mexico 87545, U.S.A.\footnote{email address:
mmn@pion.lanl.gov} \\

\vspace{0.25in}

\normalsize{ABSTRACT}

\end{center}
\begin{quotation}
\small

This article reports on a program to obtain and understand coherent states for
general systems.  Most recently  this has included
supersymmetric systems.  A byproduct of this work has been studies of
squeezed and supersqueezed states.  To obtain a physical understanding
of these systems has always been a primary goal.  In particular, in the
work on supersymmetry  an attempt
to understand the role of Grassmann numbers in quantum mechanics has been
initiated.

\end{quotation}

\section{Introduction}

In 1926, Schr\"{o}dinger published a paper [1] which described what we today
call
the ``coherent states."  This paper was separate from his fundamental
series  on ``Quantization as an Eigenvalue Problem."
Schr\"{o}dinger sent a copy of this paper to Lorentz in response to Lorentz's
objection to using  wave packets to represent particles (since the packets
must
spread out with time) [2].

{}From what we would call a knowledge of the generating function for
the Hermite polynomials, Schr\"{o}dinger was able to show that in a harmonic
oscillator potential,
a general Gaussian wave with the width of the ground state
could have arbitrary energy and momentum, follow the classical motion of a
classical particle in the potential, and not change its shape with
time. This insight eventually became formulated in one of the standard
ways to define
the coherent states (of the harmonic oscillator), as minimum uncertainty
coherent states (MUCS).

In the early 1960's, stimulated by the
classic work of Klauder [3,5], Sudarshan [4,5], and Glauber [6], there was a
renewed interest in these states in the context of quantum optics.  By using
the  boson operator techniques  of these authors, the coherent states can be
defined as displacement operator coherent states (DOCS) or as annihilation
operator coherent states (AOCS).  The displacement operator method uses
symmetry,
or group theory, techniques.

For the harmonic oscillator all three methods are equivalent, but for other
systems they are not, in general.  After reviewing these methods for the
harmonic oscillator, I will discuss extensions of them to general potentials
(MUCS) and arbitary  symmetries (DOCS). In these contexts, the place of the
``squeezed states" of the harmonic oscillator  then follows.

Although my own work for  bosonic systems has strongly emphasized the
physically intuitive minimum-uncertainty approach, when I and my colleagues
came to consider supercoherent states, we reached the conclusion that the best
approach there was to use the more abstract DOCS method.
This method provided an explicitly defined mathematical approach which we hoped
would yield physical insight into what such a bosonic-fermionic system
means, something that, {\em a priori}, was opaque.

In the final sections I will review this work for coherent states, announce
some new results for squeezed states, and discuss some physical conclusions
which can be inferred from this work.  This includes an indication of what the
role of Grassmann numbers in quantum mechanics might be.

\section{Coherent states}

	The harmonic oscillator Hamiltonian,
\begin{equation}
H = \frac{1}{2m} p^2 + \frac{1}{2}m\omega^2 x^2,
\end{equation}
is  quadratic in the operators, $x$ and $p$, which classically vary as
$\sin(\omega t)$ and $\cos(\omega t)$,  The commutation relation of the
associated
quantum operators ($\hbar = 1$)
\begin{equation}
[x,p] = i,
\end{equation}
defines an uncertainty relation
\begin{equation}
(\Delta x)^2(\Delta p)^2 \geq 1/4.
\end{equation}

\begin{quotation}

	I: (MUCS). The minimum uncertainty coherent states for the harmonic oscillator
potential can be defined as those states which minimize the uncertainty
relation (3), subject to the added constraint that the ground state is a member
of the set.

\end{quotation}

Those states which minimize the uncertainty relation (3) are
\begin{equation}
\psi(x) = [2 \pi \sigma^2]^{-1/4}
\exp\left[-\left(\frac{x-x_0}{2 \sigma}\right)^2+ip_0x\right],
\end{equation}
\begin{equation}
\sigma = s\sigma_0 = s/[2m\omega]^{1/2}.
\end{equation}
 When $s=1$, these Guassians have the width of the ground
state, so they are the coherent states.
The states are labeled by two parameters, $x_0 = \langle x\rangle $ and $p_0 =
\langle p\rangle $.

Now consider  the displacement operator approach, which was first championed by
Klauder [7].  Consider the oscillator algebra defined by $a, a^+, a^+a$, and
$I$.
The displacement operator is the unitary exponentiation of the elements of the
factor algebra, spanned by $a$ and $a^+$:
\begin{equation}
D(\alpha) = \exp[\alpha a^+ - \alpha^* a]
=\exp\left[-\frac{1}{2}|\alpha|^2\right]
\exp[\alpha a^+] \exp[-\alpha^* a],
\end{equation}
where the last equality comes from using
a Baker-Campbell-Hausdorff relation.

\begin{quotation}

II: (DOCS).  The displacement operator coherent states are obtained by applying
the displacement operator $D(\alpha)$ on  an extremal state, i.e., the
ground state.
\end{quotation}

Specifically, this yields
\begin{equation}
	D(\alpha)|0\rangle  = \exp[\alpha a^+ - \alpha^* a] |0\rangle
=\exp\left[-\frac{1}{2}|\alpha|^2\right] \sum_{n} \frac{\alpha ^n}{\sqrt{n!}}
|n\rangle
\equiv |\alpha\rangle  ,
\end{equation}
 where $|n\rangle $ are the number states.  With the identifications
\begin{equation}
Re(\alpha)=[m\omega/2]^{1/2}x_0, \hspace{.5in} Im(\alpha)=p_0/[2m\omega]^{1/2},
\end{equation}
these are the same as the MUCS.

The third definition I mention here is

\begin{quotation}
III: (AOCS).  The annihilation operator coherent states are the eigenstates of
the destruction operator:
\end{quotation}
\begin{equation}
a|\alpha\rangle  = \alpha|\alpha\rangle .
\end{equation}
(For the harmonic oscillator, III can be shown to follow from II.)

\section{Coherent states for general potentials and symmetries}

	At the end of his 1926 paper [1], Schr\"{o}dinger predicted that similar
(coherent) states could be constructed for the hydrogen atom, but that it would
be
difficult. (He never returned to the problem.)  Some 50 years later, Pete
Carruthers asked me if I thought it could be done for arbitary potentials.
In striving to answer Pete's question Mike Simmons, Vincent Gutschick, and
myself developed our generalization of the minimum-uncertainty method.

Consider an arbitary classical Hamiltonian and find those classical variables,
$X_c(x_c,p_c)$ and $P_c(x_c,p_c)$, that vary as $\sin(\omega t)$ and
$\cos(\omega
t)$.  Our first {\em ansatz} was that the classical Hamiltonian would be
quadratic
in these variables.  (For the many systems we studied, it was.)  Now turn these
classical variables into quantum operators, $X$ and $P$, by turning $x_c$ and
$p_c$ into the quantum operators $x$ and $p$ and then appropriately
symmetrizing
them in the functionals $X$ and $P$.  These operators define a commutation
relation
and hence an uncertainty relation
\begin{equation} [X,P] = iG, \hspace{0.5in}
(\Delta X)^2(\Delta P)^2 \geq \small{\frac{1}{4}}\langle G\rangle ^2.
\end{equation}
The state which minimizes the uncertainty relation (10) is given by the
solution to the eigenvalue equation
\begin{equation}
\left(X + \frac{i\langle G\rangle }{2(\Delta P)^2} P\right)\psi_{mus}
=\left(\langle X\rangle +\frac{i\langle G\rangle }{2(\Delta P)^2}\langle
P\rangle
\right)\psi_{mus}.
\end{equation}
Note that of the  four parameters $\langle X\rangle , \langle P\rangle ,
\langle P^2\rangle $, and $\langle G\rangle $, only three are
independent because they satisfy the equality in Eq. (10).

	Our second {\em ansatz} is our definition for

\begin{quotation}
I: General MUCS for Arbitrary Potentials.  The coherent states are the states
$\psi_{mus}$, subject to the restriction that the ground state  solution of
the  Schr\"{o}dinger equation be a member of the set.  (It always turned out to
work.)  This means that the three independent parameters are now reduced to
two.
\end{quotation}

As indicated, this definition worked for every solvable potential we tried.
(In WKB approximation it holds in general.) The results are  described in a
series
of papers [8], and a summary article is reprinted in the book by Klauder and
Skagerstam [9].   We even produced a 16 mm color-sound film on the time
evolution
of these coherent states. It was later made into a video [10].  (Unfortunately,
I have not found it on the shelves at any Blockbuster Video Store.)  As hoped
for, these states can be shown to follow the classical motion. They disperse
with time, as they have to, since the eigenenergies are in general not
commensurate.  The variation of decoherence time from system to system can also
be understood.  These states maintain their coherence as well as or better than
those from other methods.  In the end this is not too surprising since they
were
physically designed to do so.  It is this physically intuitive basis  for these
states which is one of their advantages.

Numerous  times John Klauder expressed an interest in this program, and his
``Doubting Thomas," penetrating questions led to specific results on at least
two
occasions:   i) the numerical comparison of the time evolution of our states
with
continuous representation states, and  ii) an explicit  demonstration of the
resolution of the identity [11].

Continuing, the generalization of the DOCS method to other symmetries is
clear.  Its application to arbitrary Lie groups has
been discussed by many people [7,12].

\begin{quotation}
II: General DOCS for Arbitrary (Lie) Symmetries.  The general DOCS are those
states
obtained by applying
the displacement operator, which is the unitary exponentiation of the elements
of
the factor algebra, on  the extremal state. This is the extremal weight
vector for noncompact groups.  (l'd call it the ground state but John would
wince.)
\end{quotation}
One important  advantage of this method is that
it presents a well-defined mathematical procedure to obtain the states.

\section{Squeezed states}

 The squeezed states of the harmonic oscillator are very easy to
obtain from the MUCS
point of view [13].  Look back to Eqs. (4-5).  Simply let $s \neq 1$, and you
have the ``squeezed states."  That is, they are minimum uncertainty Gaussians
whose widths are not necessarily that of the ground state.  This is a
continuous
three-parameter set of states.

Similarly, the generalization of squeezed states for arbitrary potentials works
the same.  These states are one step back from the general MUCS.  They are the
three parameter set of states defined in Eq. (11), $\psi_{mus}$.

The displacement operator squeezed states for the harmonic oscillator are more
complicated. Consider the ``unitary squeeze operator"
\begin{eqnarray}
S(z) & = & \exp[{\frac{1}{2}}za^+a^+ - {\frac{1}{2}}z^*aa]\\
& = &  \exp[{\small{\frac{1}{2}}}e^{i\phi}(\tanh r)a^+a^+]
\left({\frac{1}{\cosh r}}\right)^{({\small{\frac{1}{2}}+a^+a})}
\exp[-{\small{\frac{1}{2}}}e^{-i\phi}(\tanh r)aa],
\end{eqnarray}
where Eq. (13) is obtained from a BCH relation [14]. The squeezed
states equivalent to the $\psi$ of Eqs. (4-5) are obtained by operating on the
ground state by
\begin{equation}
T(\alpha,z)|0\rangle  = D(\alpha)S(z)|0\rangle  \equiv |\alpha,z\rangle ,
\end{equation}
\begin{equation}
z \equiv re^{i\phi}, \hspace{0.5in} r = \ln s.
\end{equation}
[$\phi$ is a phase which defines the starting time, $t_0 =
(\phi/2\omega)$, and $s$ is the wave-function squeeze of Eq. (5).]
But here one sees a difference.  $S(z)$ by itself can be considered to be the
displacement operator for the group SU(1,1) defined by
\begin{equation}
K_+ = \frac{1}{2}a^+a^+,\hspace{0.5in} K_- = \frac{1}{2}aa,\hspace{0.5in}
K_0 = \frac{1}{2}(a^+a + \frac{1}{2}),
\end{equation}
so that the $S(z)|0\rangle $ by themselves are the ``SU(1,1) coherent states."

It is not clear how one generalizes DOCS squeezed states to other systems,
e.g.,
to the $\cosh^{-2}$ potential.   (Obviously, squeezed states
represent a  more complicated symmetry than coherent states.)  But the MUCS
generalization is clear.

Bob Fisher, Vern Sandberg, and myself looked at the possiblity of
naively generalizing the harmonic oscillator squeezed states to higher order in
$a$
and $a^+$ [15].  It turned out this could not be done. (Once again John's hand
crept in, helping us along the way.)  This background [13-15] was later an aid
to
Alan Kosteleck\'{y}, Rod Truax, and myself, when we came to the problem of
``supersqueezed states."

\section{Supercoherent states}

	The introduction of
graded Lie algebras was an important milestone in the study of
combined internal and space-time symmetries.  This led to the
development of supersymmetric theories which predict the existence
of boson and fermion partner states:  e.g., for the photon there
is a partner photino, etc.

	Up to now none of the
supersymmetric partners have been found, so that supersymmetry, if
extant, is broken.  All evidence for supersymmetry has been found
in the low energy regime: e.g., in nuclei [16] in atomic systems [17] and in
WKB calculations [18]. But these results, although exciting and indicative,
do not unequivocally prove the need for a fundamental supersymmetry. Rather,
they
are tantalizing hints.  The proof would be in the discovery of supersymmetric
particles, perhaps at the SSC.

Our own interest in physical manifestations of supersymmetry in atomic
systems [17], combined with our interest in coherent states, led
Kosteleck\'{y}, Truax, and myself to consider how one should define
``supercoherent states."  We came to the conclusion that we should not seek to
generalize the minimum-uncertainty method, because just how that should
be applied to the fermion sector was unclear.

In the MUCS program,
physical intuition had led to the mathematics.  But here we felt we should let
the
mathematics lead us to the physical intuition.  That is to say, by using the
displacement operator method, but with supergroups, we had a
method that was  well-defined and which we could use  because we had already
developed the theory and application of super-BCH relations [19].  This gave us
a
tool that other workers in the field did not have [20].

Joined by Beata Fatyga, Alan's graduate student, we derived our general
supercoherent states [21].  (As before, John Klauder could be found giving
advice.)  The definition is the same as the DOCS method for general Lie groups
given  above, only one uses supergroups and their associated (anti)commutation
relations.  We presented three examples, with physical models for each: i) The
super oscillator algebra defines the supersymmetric harmonic oscillator. ii) A
supersymmetric quantum-mechanical system was given which has a
Heisenberg-Weyl albegra plus another bosonic degree of freedom.  This
represents
an electron in a constant magnetic field. iii) An OSP(1/2) supersymmetry
represents the electron-monopole system.

Of the three examples, the first is the simplest and 	I use it to demonstrate
the
technique.  The super oscillator algebra is defined by
\begin{equation}
	[a,a^+] = I,  \ \ \   \{b,b^+ \} = I   .
\end{equation}
Using  super-BCH relations, one obtains that the super displacement operator is
\begin{eqnarray}
{\bf D}(g) & = & \exp[ Aa^+ -\overline{A}a + \theta b^+ + \overline{\theta}b]
\\	 & = & \left(\exp[-\frac{1}{2}|A|^2]\exp[A a^+] \exp[-\overline{A} a]\right)
\left(\exp[-\frac{1}{2}\overline{\theta}\theta]\exp[\theta b^+]
\exp[\overline{\theta} b]\right)\\
& \equiv & D_B(A)\,D_F(\theta).
\end{eqnarray}
The $B$ and $F$ subscripts denote the fact that our supersymmetric
displacement operator can be written as a product of ``boson" and ``fermion"
(more
properly, even and odd) displacement operators.

$\theta$ and $\overline{\theta}$ are odd Grassmann numbers. They  are nilpotent
(they only contain a ``soul"). They satisfy anticommutation relations among
themselves and with the fermion operators $b$ and $b^+$.  $A$ and
$\overline{A}$
are complex, even, Grassmann numbers.  Because $a^+$ and $a$ are pure even
elements, we associate the ``body" of $A$ and $\overline{A}$ with the $\alpha$
and
$\alpha^*$ of the ordinary coherent states.  The ``soul" part of $A$ must be
even in the Grassmann numbers.  For example,
$A$ could be of the form
\begin{equation}
 A = \alpha + \tau\overline{\theta}\theta,
\end{equation}
\noindent although  technically $A$ is not
restricted to the two odd basis elements we have.

	 Explicit calculation yields
\begin{equation}
	D_S(g)|0,0\rangle
	= [1 - (1/2)\overline{\theta}\theta ]|A,0\rangle  + \theta |A,1\rangle
\equiv |Z\rangle   .
\end{equation}
The two labels of
$|0,0\rangle $in Eq. (22) represent the even (bosonic) and odd (fermionic)
sectors.  The bosonic sector has the form of an ordinary  coherent state
$|A\rangle $ and the fermionic sector has zero or one fermions.  (I refer the
reader to Ref. [21] for further details.)  If the bosonic displacement is
turned
off, then the fermionic displacement produces states with zero and one fermion,
but no bosons.   (In Sec. 7, I will return to an idea on how to  physically
interpret these states.)

\section{Supersqueezed states}

{}From the previous sections, it is clear that the supersymmetric
generalization of
the SU(1,1) squeeze operator of Eqs. (12-13) is what is desired.  The group
involved is the supergroup OSP(2/2).  In addition to the SU(1,1) elements of
Eq.
(16), it has five more:
\begin{eqnarray}
M_0 &=& \frac{1}{2}(b^+b-\frac{1}{2}), \nonumber \\
Q_1&=& \frac{1}{2}a^+b^+,  \hspace{.3in}  Q_2 = \frac{1}{2}ab, \hspace{.3in}
Q_3 = \frac{1}{2}a^+b,    \hspace{.3in}  Q_4 = \frac{1}{2}ab^+.
\end{eqnarray}

The commutation relations follow, and so the supersqueeze operator can in
principle be written as ($\hat{g}$ is the factor algebra)
\begin{eqnarray}
{\bf S}(g) & = & \exp\left[\sum_{i=1}^{6}\alpha_i \hat{g}_i \right] \\
& = &\prod_{i=1}^{8}\exp[\beta_i g_i].
\end{eqnarray}
The above can be solved by using super-BCH relations. They yield twenty
coupled differential equations.  (The even operators
have separate equations for order zero, two and four  Grassmann numbers, and
the odd operators have separate equations for orders one and three.)

We have just finished solving these equations and are in the process of
performing some final calculations [22].  For now it is useful to note that the
squeeze operator can again  be separated into a product of bosonic and
fermionic
pieces: \begin{equation}
{\bf S}(g) = S_B(g)\,S_F(g) .
\end{equation}
The fermionic squeeze operating by itself on $|0,0\rangle$  produces the
states $|0,0\rangle ,|1,1\rangle $, and $|2,0\rangle $.

Therefore, one finally obtains that the supersqueezed states are, in general,
of
the form \begin{eqnarray}
{\bf T}(g) |0,0\rangle  &=& {\bf D}{\bf S}|0,0\rangle  \nonumber \\
&=&D_B(A)D_F(\theta)S_B(g)S_F(g)|0,0\rangle
   = [D_B(A)S_B(g)][D_F(\theta)S_F(g)]|0,0\rangle  \nonumber \\
&\equiv& {\bf T}_B(g){\bf T}_F(g)|0,0\rangle .
\end{eqnarray}
The general  operator produces states with arbitary numbers of bosons
and zero or one fermion.

\section{Discussion}

If supersymmetric partners were actually found, then
fundamentally one would need to
give a physical interpretation to  Grassmann numbers.  This  situation would be
similar to the problem of the physical interpretation of imaginary numbers when
quantum mechanics was discovered.   When
Schr\"{o}dinger described his ``coherent states" [1], he thought that
the physics was contained in the real part of his wave solutions.   The
realization that the complex phase had physical information came later.

Recently I addressed this problem [23], although I was by no means the first
to do so [24].    	Looking at the
supercoherent states of Eq. (22),  I suggested that the odd part
describes the existence of a coherent, massless fermion; i.e., a
"photino" with energy $E$, it being coherent with the various  n-photon
states in the bosonic sector.

As to the
Grassmann numbers, I made the  suggestion that the fermion sector ``phase"
relative to the boson sector is defined by the $c$ in $\theta \equiv c\zeta$,
$\zeta$ being a Grassmann basis vector and labeling the fermionic part of the
state.  The probability of finding a supercoherent state that has a bosonic
sector
coherent with one photino is $c^* c$, with the $\overline{\zeta}\zeta$ in
$\overline{\theta}\theta$ labeling the probability as being for a fermion.
Thus, the probability of finding a bosonic sector without a coherent photino is
$(1 - c^* c)$, from $(1 - \overline{\theta}\theta)$.  Note that one must have
$|c| \leq 1$.  Then the probabilities for the coherent state having or not
having
a fermion are both $\leq 1$.

The restriction on
the value of $c$ is analogous to physical restrictions placed on ordinary
quantum mechanics.  For example, one demands that all state probabilities
$P_n = |a_n|^2 \leq 1$ and one disallows unnormalizable solutions of the
Schr\"{o}dinger equation.  (See Ref. [23] for more details.)  Even so, I must
emphasize that at this point this restriction on $c$ is based on physical
intuition
rather than on mathematical rigor.

Now that we have our supersqueezed states [22], I hope to pursue this line of
thought, taking into account the insight available from studying the Grassmann
structure of the supersqueezed states.  It appears that a similar restriction
can be made for the quantities analogous to $c$ in the supersqueezed states.
Just as as was the case for Ref. [23], I expect useful comments to come
from John Klauder.\\

{\large{\bf Acknowledgements}}\\

Of course, in a review of a program such as this, I must once again acknowledge
my colleagues and friends with whom I have enjoyed working over the years.

But most all, this is John Klauder's celebration.   As the reader will have
noted,
the influence of John Klauder can be seen throughout this work.  It is an
honor, but most of all it is a great personal pleasure, to congratulate John on
the
occasion of his 60th birthday.  I eagerly await our next discussion, perhaps on
Grassmann numbers and supersqueezed states, when I can once again  hear John
say,``But Michael,\ldots"

\end{document}